\documentclass[12pt]{article}
\usepackage{epsfig}

\begin{document}

\begin{titlepage}
\title{A new approach to the study of the ground-state\\
 properties of 2D Ising spin glass}
\author{ Zhi Fang Zhan, Lik Wee Lee, Jian-Sheng Wang \\
         Department of Computational Science,\\ 
         National University of Singapore,\\ 
         Singapore 119260, Republic of Singapore}

\maketitle
\thispagestyle{empty}

\begin{abstract}
   
A new approach known as flat histogram method is used to study the $\pm J$ 
Ising spin glass in two dimensions. Temperature dependence of the energy, 
the entropy, and other physical quantities can be easily calculated and we
give the results for the zero-temperature limit. For the  
ground-state energy and entropy of an infinite system size, we estimate  
$e^0 = -1.4007 \pm 0.0085$ and $s^0 = 0.0709 \pm 0.006$, respectively. 
Both of them agree well with previous calculations. 
The time to find the ground-states as well as the tunneling times of the 
algorithm are also reported and compared with other methods. 

                 \\  

\end{abstract}


\end{titlepage}

\section{Introduction}
\par The equilibrium properties of spin glass have remained a
great challenge in numerical simulations. Investigating the equilibrium
ground-state structure of spin glass is also important and interesting. 
In the last 20 years, there has been a great deal of work on spin
glass~\cite{BinderYoung}. 
It is generally agreed that the simplest spin glass system for most theoretical
work is the Edwards-Anderson (EA) model, whose Hamiltonian is     
\begin{equation}
	H = -\sum_{<i,j>} J_{ij} \sigma_i \sigma_j,
\end{equation}
where $\sigma_i$ takes on the values $\pm 1$ and the sum goes over 
the nearest neighbors. The $J_{ij}$ are dimensionless variables
which describe the random interactions between the spins and are 
taken as $J_{ij} = \pm 1$. In two dimensions,
a phase transition occurs only at zero 
temperature~\cite{Mor,McMillan,SwenWang1} for this
kind of $\pm J$ Ising spin glass with nearest neighbor interactions. 
This model has been studied previously by the transfer matrix 
method~\cite{Mor,Cheung}, replica Monte Carlo 
method~\cite{SwenWang1,SwenWang2}, multicanonical ensemble 
method~\cite{Berg1} and many other methods (see ref~\cite{BinderYoung}
for a review).
\par The traditional Monte Carlo methods mostly concentrate on generating 
standard statistical ensembles, e.g., the canonical ensemble or
microcanonical ensemble. Using the canonical ensemble simulations, we need
to simulate at different temperatures to get full information
about the system. It is tedious to calculate certain thermodynamic quantities 
like the free energy and the entropy since the density of states cannot be 
obtained directly from the simulation data.  
The correlation between subsequent configurations generated by canonical
ensemble simulations also causes the ergodicity problem for some systems.
In 1991, Berg proposed the multicanonical ensemble method~\cite{Berg2} 
to overcome the above shortcomings of simulations on canonical 
ensemble. The multicanonical ensemble is an ensemble where the probability
$P(E)$ of having energy $E$ at equilibrium is a constant. 
The multicanonical method has been very successful in solving the systems
that involve energy barriers.
\par Recently, Wang proposed a dynamics~\cite{Wang1} which can generate a 
flat histogram in the energy space as the multicanonical method.
This dynamics has some connections with the broad histogram 
method~\cite{Broad}, which does not give the correct microcanonical 
average~\cite{Wang1}. Similar to the broad histogram method,
the new dynamics is also based on $\langle N(\sigma,\Delta E) \rangle$, the
(microcanonical) average number of potential moves which increase
the energy by $\Delta E$ in a single spin flip. A cumulative average (over 
Monte Carlo steps) can be used as a first approximation to
the exact microcanonical average in the flip rate. Thermodynamic 
quantities can be then calculated from the simulation data with ease. 
In this paper, we use the new method to study the thermodynamics 
as well as ground-state properties for the two-dimensional Ising spin
glass system. 
\par In Section 2, the flat histogram transition matrix Monte Carlo dynamics
is described. Using the flat histogram sampling, we get the average number of 
potential moves $\langle N(\sigma, \Delta E) \rangle _E$, which can be used to 
construct a transition matrix Monte Carlo 
dynamics in the energy space~\cite{Wang2}. We apply the new method to 
two-dimensional Ising spin glass and present some numerical results in 
Section 3. In the last section, we give a conclusion to the new method. 

\section{The transition matrix Monte Carlo dynamics with the flat 
histogram sampling}
To connect our dynamics with single-spin-flip Glauber dynamics~\cite{Glauber}, 
we restrict the protocol of each move to be single-spin flip in the 
following discussion. For a given state $\sigma$ with energy $E$, consider all 
possible single-spin flips. The single-spin flips change the current state 
into $N$ possible new states, with new energy $E' = E + \Delta E$. 
For two-dimensional Ising spin glass, $\Delta E$ = $0$, 
$\pm 4$, and $\pm 8$. We classify the $N$ new states according to $\Delta E$ and
count the number of $N(\sigma,\Delta E)$.   
Since each move from the state $\sigma$ of energy $E$ to the state
$\sigma'$ of energy $E'$ and the reverse move are both allowed, the total
number of moves from all the states with energy $E$ to $E'$ is the same as 
from $E'$ to $E$. Thus, we have~\cite{Wang3}
\begin{equation}\label{balance}
\sum_{E(\sigma) = E} N(\sigma,\Delta E) = 
\sum_{E(\sigma') = E+\Delta E} N(\sigma', -\Delta E).
\end{equation}
The microcanonical average of a quantity $A(\sigma)$ is defined as
\begin{equation}
\langle A \rangle _E = \frac{1}{n(E)} \sum_{E(\sigma) = E} A(\sigma),
\end{equation}
where the summation is over all the configurations having energy $E$ and
$n(E)$ is the density of states.
In terms of the microcanonical averages, we can rewrite Eq.~(\ref{balance}) as 
\begin{equation}\label{detail}
 n(E) \langle N(\sigma,\Delta E) \rangle_E =
 n(E+\Delta E) \langle N(\sigma',-\Delta E) \rangle _{E+\Delta E}.
\end{equation} 
Eq.~(\ref{detail}) is the basic result 
of the broad histogram method~\cite{Broad}. While the broad histogram random
walk algorithm is not correct, Eq.~(\ref{detail}) is not problematic and taken
as the starting point of the flat histogram sampling.
\par We select a site to flip at random. The flip rate for a single-spin 
flip from state $\sigma$ with energy $E$ to $\sigma'$ with energy  
$E' = E + \Delta E$ is chosen as
\begin{equation}
r(E'|E) = \min \Big(1,\frac{\langle N(\sigma',-\Delta E) \rangle _{E'}} 
{\langle N(\sigma,\Delta E) \rangle _E} \Big).
\end{equation}
Then the detailed balance condition for this rate   
\begin{equation}
r(E'|E)P(\sigma) = r(E|E')P(\sigma')
\end{equation}
is satisfied for $P(\sigma) \propto 1/n(E(\sigma))$. Thus the energy 
histogram is flat~\cite{Wang3}, 
\begin{equation}
P(E) = \sum_{E(\sigma) = E} P(\sigma) \propto n(E) \frac{1}{n(E)} = 
\textnormal{const}.
\end{equation}
\par Since $\langle N(\sigma,\Delta E) \rangle _E$ is not known in general, 
an approximation scheme
should be used to start the simulation. For those $E$ which we have not
visited yet, we simply set $r(E'|E) = 1$. Then a cumulative average (over 
Monte Carlo steps) can be used as an approximation to the exact 
microcanonical average in the flip rate. We have numerical evidence that
this procedure converges to the exact result.
\par We can then construct a transition matrix Monte Carlo dynamics in the 
energy space~\cite{Wang2} with $\langle N(\sigma,\Delta E) \rangle _E$.
For a single-spin-flip Glauber dynamics with 
energy change $\Delta E$, the flip rate is given as
\begin{equation}
w(\Delta E) = \frac{1}{2} \Big[1-\tanh\Big(\frac{\Delta E} {2k_BT}\Big) \Big] .
\end{equation}
Since there are (on average) $\langle N(\sigma,\Delta E) \rangle _E$ 
different ways of going from $E$ to $E' = E + \Delta E$, the total
probability for transition from $E$ to $E'$ is
\begin{equation}
W(E + \Delta E|E) = w(\Delta E) \langle N(\sigma,\Delta E) \rangle_E, \qquad 
for~\Delta E \ne 0 \label{weight}.
\end{equation}
The diagonal elements can be determined by 
$\sum_{\Delta E} W(E + \Delta E|E) = 1$,
since the total probability from $E$ to $E'$ is 1.
This new dynamics in the space of energy $E$ is related to 
single-spin-flip dynamics by~\cite{Wang2}
\begin{equation}
W(E'|E) = \frac{1}{n(E)}\sum_{E(\sigma) = E}\sum_{E(\sigma') = E'} \Gamma(
\sigma'|\sigma).
\end{equation}
where $\Gamma(\sigma'|\sigma)$ is the transition matrix of the 
single-spin-flip dynamics. 
The equilibrium state of the transition matrix gives
the canonical probability distribution of energy 
$P_T(E) \propto n(E) \exp(-E/k_BT)$.
\par An important aspect of this dynamics is that
we can calculate the thermodynamic quantities easily by just performing one
simulation for each coupling state $J_{ij}$. The density of states
$n(E)$ can be obtained through Eq.~(\ref{detail}). Once we have the
density of states $n(E)$, we can obtain $P_T(E)$ and then calculate any 
thermodynamic quantities of interest. In actual implementation, we usually 
determine $P_T(E)$ directly from the detailed balance equation 
\begin{equation}
W(E+\Delta E|E)P_T(E) = W(E|E+\Delta E)P_T(E+\Delta E)
\end{equation} 
instead of solving Eq.~(\ref{detail}).
From Eq.~(\ref{weight}), we know, the transition matrix $W(E'|E)$ can
be formed at any temperature once the 
quantity $\langle N(\sigma,\Delta E) \rangle _E$ is
computed accurately. In other words, the Monte Carlo computation is 
uncorrelated to thermodynamics. The temperature dependence enters only after
simulation in the weighting formula. 
\par Like Berg's multicanonical ensemble simulations, our dynamics also 
generate a multicanonical ensemble in the energy space. From this point,
both of the two dynamics have the same goal of flattening the space of
energy. But they are quite different in implementation. 
In the multicanonical ensemble method, the flip rate is chosen as the inverse 
of the density of states $n(E)$, parametrized in some way. To start
the simulation, we need give an estimate of $n(E)$, since $n(E)$ is not
initially known. Thus, the efficiency of this method is determined by the
goodness of the estimated $n(E)$. If $n(E)$ is not given properly, say far
off the true density, the simulations may get stuck in some region. With our
method, we sample the energy space with a flip 
rate which is related to the density of states through Eq.~(\ref{detail}). 
The central quantity is $\langle N(\sigma, \Delta E) \rangle _E$ which
can be quite accurate in a short simulation time. And the accuracy 
of this quantity is improved by further simulations. We then provide
an alternate for the estimate of $n(E)$, which leads to a
more efficient way for simulating the multicanonical ensemble.\\ 

The flat histogram also generalizes easily to multi-variate models \cite{Lima}.
An example is the Ising Spin Glass model with overlap parameter $q$ which has
the Hamiltonian
\begin{equation}
H_2 = -\sum_{<i,j>} J_{ij}\sigma_i^{1}\sigma_j^{1}
-\sum_{<i,j>} J_{ij}\sigma_i^{2}\sigma_j^{2}
-h\underbrace{\sum_i\sigma_i^1\sigma_i^2}_{q}.
\end{equation}
$E$ refers to the first two interaction terms involving the coupling constants
$J_{ij}$.
In this bivariate case, the quantity $\langle N(\sigma, \Delta E) \rangle$
generalize to $\langle N(\sigma, \Delta E, \Delta q) \rangle$. It can be
easily shown that the detailed balance condition is now
\begin{equation}
n(E,q) \langle N(\sigma, \Delta E, \Delta q) \rangle_{E,q} =
n(E+\Delta E, q+\Delta q) \langle N(\sigma', -\Delta E, -\Delta q)
\rangle_{E+\Delta E, q+\Delta q}
\end{equation}
with $n(E,q)$ as the new ``density of states". The algorithm gives a flat
histogram in both $E$ and $q$.

\section{Numerical results}
We have performed simulations on lattices of size $L = 4, 10, 16, 24$ and $32$. 
Each simulation starts with independent random numbers.
To illustrate the performance of our algorithm, we define the time
$\tau_L$ as the average number (over coupling constant $J_{ij}$) of Monte 
Carlo steps needed to reach the ground-states. A Monte Carlo step is defined 
as flipping each spin on the lattice once (on the average). 
Table 1 gives an overview of typical time in Monte Carlo steps to reach 
the ground-states, starting from an arbitrary energy level. 
The time to reach the ground-states depends on the size of the system and also
the random interactions. We consider a large number of random coupling states
to make the statistical error small enough in Table~1. The simulations are
long enough to ensure that the ground states are really reached.
In Fig.1 we plot the time $\tau_L$ versus lattice size $L$ on a double
log scale. The data are consistent with a straight-line fit, which
gives the finite-size behavior
\begin{equation}
\tau_L \propto L^{4.71}, \quad \textnormal{MC~steps}.
\end{equation}
The corresponding CPU time for a Digital Alpha 600M workstation is also shown
in Table~1. For accuracy, 5 independent runs are performed for each lattice
size to obtain the average CPU time. Up to $L = 32$ the CPU time can be
approximated by a polynomial function of $L^{6.08}$.\\
\begin{table}[ht]
\begin{center}
\begin{tabular}{p{2cm}  p{4cm}  p{4cm}}
\hline
\hline
 L  & $\tau_L$(MC Steps)  & CPU time(second)\\
\hline
 4  & 45.33 $\pm$ 0.91		& 0.003 $\pm$ 0.001\\
 10 & 2752 $\pm$ 127		& 0.43 $\pm$ 0.012\\
 16 & 25761 $\pm$ 1504		& 8.55 $\pm$ 0.55\\
 24 & 216884 $\pm$ 18444	& 189 $\pm$ 15\\
 32 & 759609 $\pm$ 80310	& 733 $\pm$ 86\\
\hline
\hline
\end{tabular}
\caption{Average MC steps and CPU time to find the ground-states for different
lattice size.}
\end{center}
\end{table}
\par We also consider the tunneling time which is defined as the average
Monte Carlo steps needed to move from $E_{max}$ to $E_{min}$, or from
$E_{min}$ to $E_{max}$. Note that $E_{min}$ is the same as ground state
energy and $E_{max}$ is $N-E_{min}$.
We note that Berg's definition about tunneling
time is slightly different from ours. During the simulation, Berg imposed
a constraint $\sum_{ij} J_{ij} = 0$. But for our method, both the time
$\tau_L$ and the tunneling time will not be affected significantly by the
imposition of the constraint. We start the simulations from an arbitrary 
energy level. Table 2 gives an overview of the tunneling time obtained using 
the two methods. The power law fits are
\begin{equation}
\tau_{M.C.} \propto L^{4.43}, \;\;\; \textnormal{and} \;\;\;
\tau_{F.H.} \propto L^{5.03},
\end{equation}
for Berg's method and current flat histogram method, respectively.
It shows that they basically give the same tunneling time.
\begin{table}[ht]
\begin{center}
\begin{tabular}{p{1.5cm} p{3cm} p{3.2cm} p{3.2cm}}
\hline
\hline
$L$ & Flat Histogram\newline (F.H.)& Multicanonical \newline (M.C.) & Bivariate  in $q$ \newline ($q$)\\
\hline
4   &   27.4 $\pm$ 0.14          & 35.3 $\pm$ 2.8	   		& 75.1 $\pm$	3.0\\
8   &   541 $\pm$ 9              & 				& ($1.82\pm0.23)\times10^3$\\
12  &   $(4.71\pm0.3)\times10^3$ & $(2.61\pm0.45)\times10^3$	& ($9.68\pm1.08)\times10^3$\\
16  &			         &				& ($3.74\pm0.31)\times10^4$\\
24  &   $(2.22\pm0.6)\times10^5$ & $(1.94\pm0.44)\times10^5$ 	&\\
48  &   		       & $(1.46\pm0.52)\times10^6$ 	&\\
\hline
\hline
\end{tabular}
\caption{Average tunneling time obtained with two dynamics for different 
lattice sizes. The multicanonical results are obtained from Ref.~\cite{Berg1}.}
\end{center}
\end{table}   

\par We also compared with Hatano's result \cite{Hatano} that autocorrelation
time scales approximately as volume $N$ of the system. We look at the tunneling
time which is a better measure of the algorithm's efficiency in our case.
From our results given in Table 2, we found no support for Hatano's result.
Instead, the power law fit (see Fig.~2)
\begin{equation}
\tau_q \propto L^{4.45}
\end{equation}
is almost the same as the monovariate case. This is not surprising as
Berg mentioned that the optimal performance for multicanonical algorithm
is $\propto N$(=$L^2$) based on random walk picture.
\par The ground-state energy and entropy of the infinite system are also
estimated using our method. It is straightforward to obtain the ground-state
energy in the simulation stage. We calculate the ground-state entropy from
\begin{equation}
S(E) = \frac{k_B}{N} \ln n(E).
\end{equation}
Since $n(E)$ can be calculated from the simulation data directly, we then
obtain $S(E)$ with ease.

To compare with the results obtained in the literature, we fit our data
using the form $f_L = f_{\infty} + c/L^2$ and get
$e^0 = -1.4007 \pm 0.0085$, $s^0 = 0.0709 \pm 0.006$. The energy fit is
plotted in Fig.~3, and the entropy fit in Fig.~4.
Our energy estimate $e^0 = -1.4007 \pm 0.0085$ is consistent with the previous
MC estimate~\cite{SwenWang1} $e^0 = -1.407 \pm 0.008$ as well as with the
transfer matrix result~\cite{Cheung} $e^0 = -1.4024 \pm 0.0012$.
Our entropy estimate $s^0 = 0.0709 \pm 0.006$ is also consistent with the
MC estimate~\cite{SwenWang1} $s^0 = 0.071 \pm 0.007$ as well as the 
transfer matrix result~\cite{Cheung} $s^0 = 0.0701 \pm 0.005$. 
For the two-dimensional Ising spin glass system, De Simone et al.~\cite{Simone}
use an exact algorithm based on the branch-and-cut technique to find
the exact ground-states with system size up to $50\times50$. They
obtain the extrapolated result $e^0 = -1.4022 \pm 0.0003$.
When compared with Berg's result, $e^0 = -1.394 \pm 0.007, 
s^0 = 0.081 \pm 0.004$, it seems that our method gives a more accurate estimate
for ground-state energy and entropy for an infinite system.
\begin{figure}
\begin{center}
\mbox{\epsfig{file=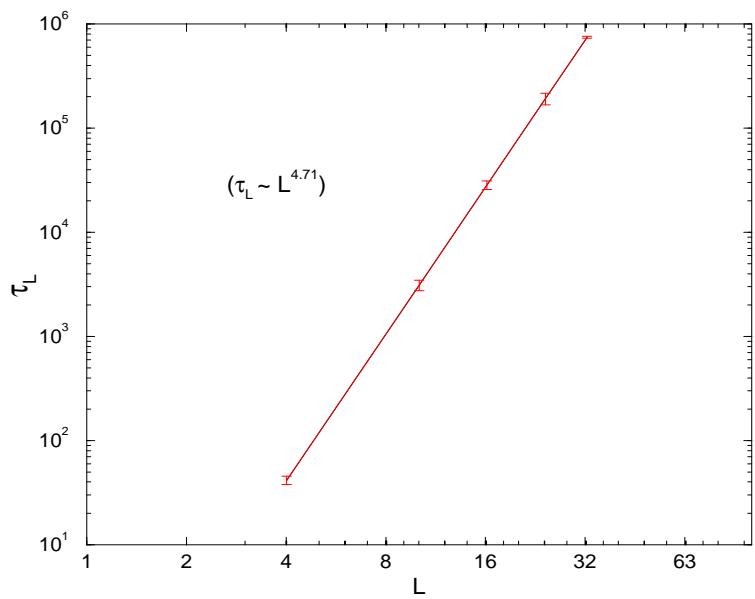, height = 8cm, width=10cm}}
\caption{$\tau_L$ vs lattice size on a double log scale.}
\end{center}
\end{figure}

\begin{figure}
\begin{center}
\mbox{\epsfig{file=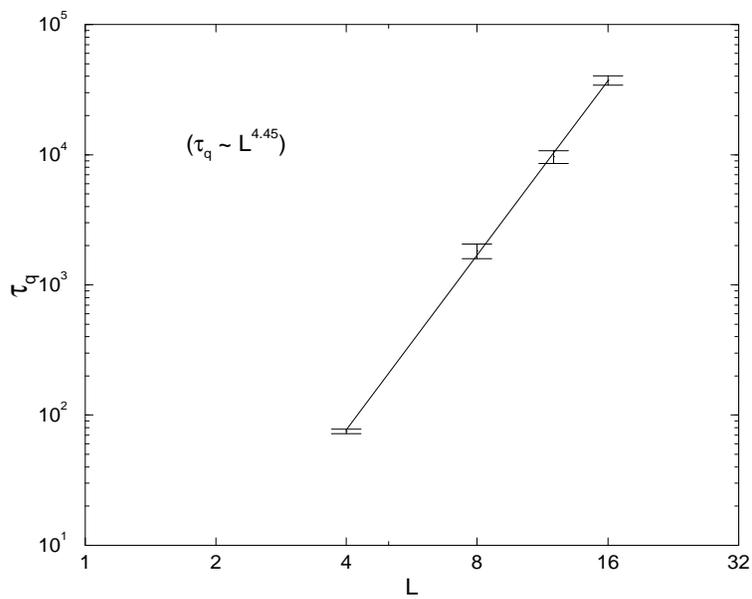, height = 8cm, width=10cm}}
\caption{Tunneling time of bivariate model on a double log scale.}
\end{center}
\end{figure}
\begin{figure}
\begin{center}
\mbox{\epsfig{file=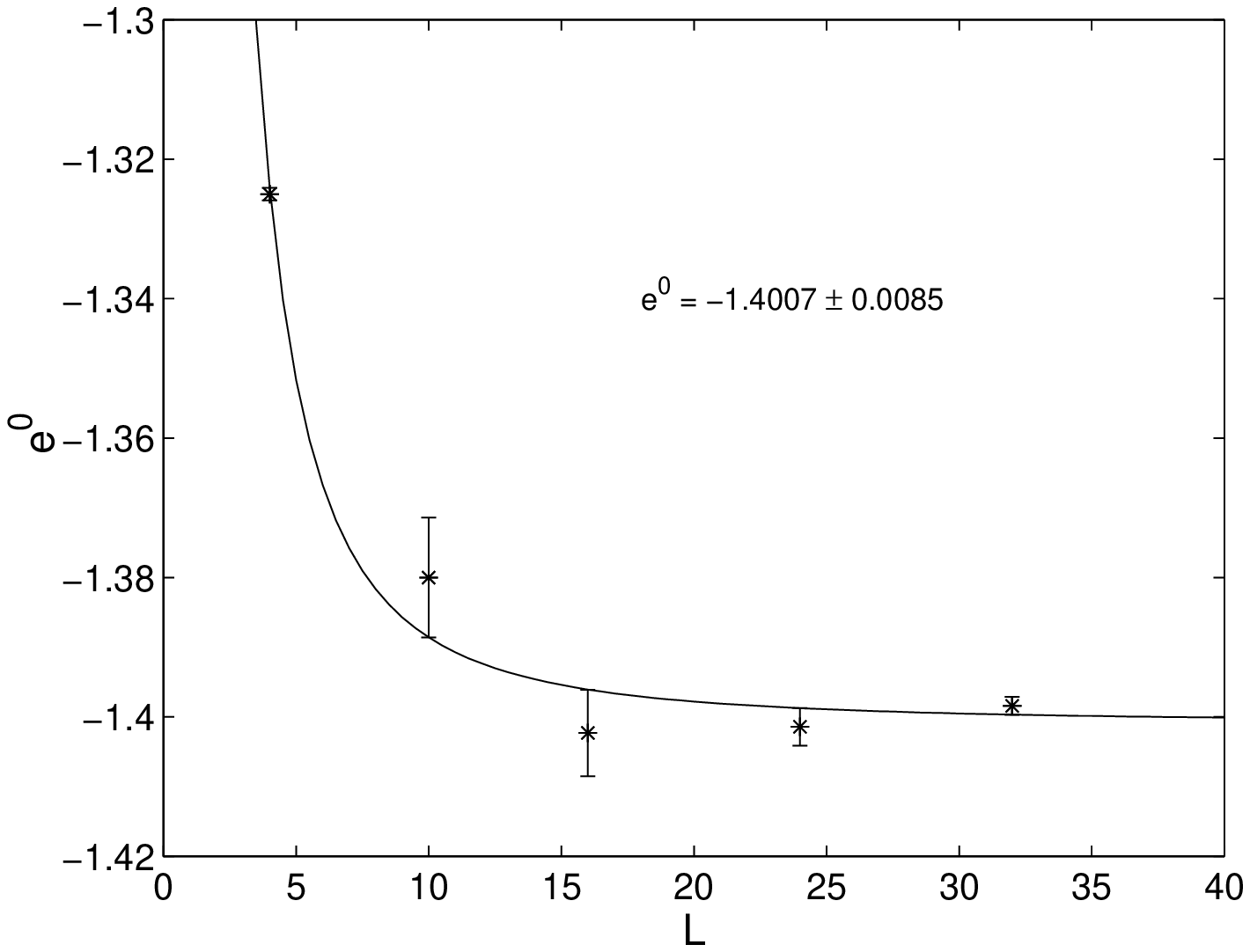, height = 8cm, width=10cm}}
\caption{FSS estimate of energy per spin of the 
infinite system size.}
\end{center}
\end{figure}

\begin{figure}
\begin{center}
\mbox{\epsfig{file=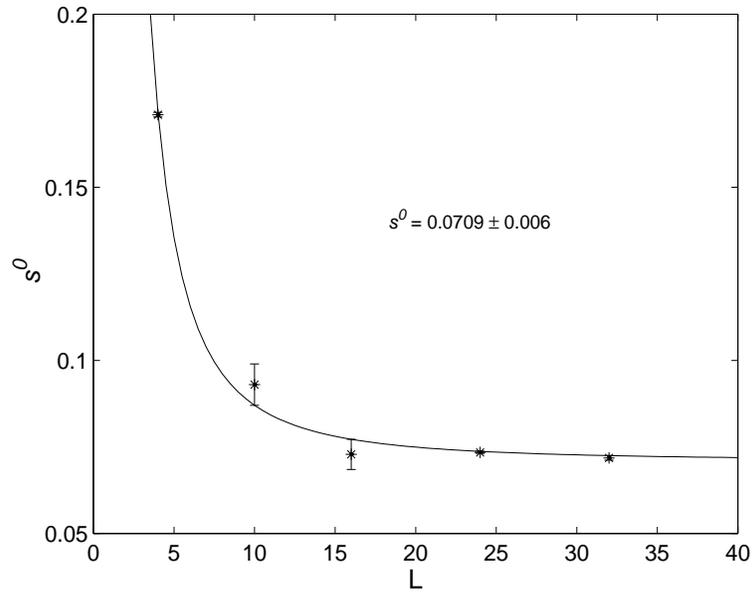, height = 8cm, width=10cm}}
\caption{FSS estimate of entropy per spin of the 
infinite system size.}
\end{center}
\end{figure}

\section{Conclusions}  
We have used a new approach to investigate the ground-state properties of 
the two-dimensional Ising spin glass. Compared with standard simulations, 
the advantage of our method is obvious.  
For the ergodicity problem encountered in standard simulations,
our method behaves as well as Berg's multicanonical ensemble method, 
while it is easier to be implemented compared with Berg's method. Our method
also generalize straightforwardly to multi-variate models without much effort
in programming and theory.
\par To find a true ground-state, we roughly need a CPU time of order $L^6$.
It is the same with Lawler's exact algorithm~\cite{Lawler}.
Up to size $50 \times 50$, De Simone's algorithm also needs a time of 
order $L^6$. But it is not clear whether his algorithm can be efficiently
implemented for 3D systems. However our method can also be easily
applied to 3D spin glass system. If one is just interested in finding 
the ground-states, there are also other optimized algorithms.
Chen's learning algorithm~\cite{Chen} is fast in finding the ground-states
compared with most algorithms, but it is not a general one.
Thermodynamic quantities cannot be obtained with this algorithm.
\par We believe that the approach we present in this paper is useful in 
studying the thermodynamics as well as ground-state properties for spin 
glass systems. 
It also can be applied to other models because of its generality.  

\end{document}